\documentstyle[aps,prl,epsf,twocolumn,floats]{revtex}
\begin{document}
\draft

%2col
\twocolumn[\hsize\textwidth\columnwidth\hsize\csname @twocolumnfalse\endcsname
% start of wide text
%2col

\title{Projecting the Kondo Effect: Theory of the Quantum Mirage}
\bigskip
\author{Oded Agam and Avraham Schiller}
\address{ Racah Institute of Physics, The Hebrew University,
                Jerusalem 91904, Israel}
%\date{\today}
\maketitle

\begin{abstract}
A microscopic theory is developed for the projection
(quantum mirage) of the Kondo resonance from one focus
of an elliptic quantum corral to the other focus. The
quantum mirage is shown to be independent of the
size and the shape of the ellipse, and experiences
$\lambda_F/4$ oscillations ($\lambda_F$ is the surface-band
Fermi wavelength) with an increasing semimajor axis
length. We predict an oscillatory behavior of the
mirage as a function of a weak magnetic field applied
perpendicular to the sample.
\end{abstract}

]
\narrowtext

In a recent experiment, Manoharan {\em et al.}~\cite{Mirage_00}
used an elliptic quantum corral to project the image of a Kondo resonance
over a distance of tens of angstroms, from one focus of the ellipse to the
other focus. By placing a magnetic Co atom at one focus of the ellipse
and measuring the tunneling current to a close-by scanning tunneling
microscope (STM) tip, a distinctive Kondo resonance was seen in the
\mbox{$I$-$V$} curve when the tip was brought directly above the Co adatom.
Remarkably, a similar Kondo signature was observed when the tip was
placed above the empty focus, indicating coherent refocusing of the
spectral image by the surrounding corral. This should be contrasted
with STM measurements of isolated magnetic adatoms on open
surfaces~\cite{STM98_ce,STM98_co}, where a limited spatial extent
of $\sim 10$\AA \ was observed for the Kondo effect.

Semiclassically, one can attribute this refocusing phenomena to the
property that all classical paths leaving one focus of the ellipse
bounce specularly off the perimeter and converge onto the second
focus with the same acquired phase~\cite{Mirage_00} (see Fig.~1).
However, this simple picture does not explain the quantitative 
features of the experiment. For example, the complex interference
patterns in the $dI/dV$ difference map throughout the ellipse, or
the $\lambda_F/4$ oscillations of the mirage with an increasing
semimajor axis length $a$ ($\lambda_F$ is the Fermi
wavelength). Explanation of these features requires a quantitative
theory, which is the objective of the present Letter.

Starting with a microscopic picture of Kondo scattering off the
Co adatoms
we obtain good qualitative and quantitative
agreement with the experiment. We establish a remarkable feature
of the quantum mirage which, aside from the $\lambda_F/4$
oscillations mentioned above, is independent of the size and the
shape of the ellipse, provided the ellipses is not too small.
In particular, there is no dependence on the
ellipse eccentricity ${\cal E}$, see Fig.~1. In the presence
of a weak perpendicular magnetic field, we predict an oscillatory 
behavior of the quantum mirage as a function of the magnetic flux
encircled by the ellipse.

The Cu(111) surface has a band of surface states, which acts
as a two-dimensional electron gas. The surface band
starts $450{\rm meV}$ below the Fermi energy, and has a Fermi
wave number of $k_F^{-1} \simeq 4.75$\AA. When a
Co adatom is placed on the surface, it scatters both the surface
electrons and the underlying bulk electrons. As recently shown
by \'Ujs\'aghy {\em et al.}~\cite{Zawa_2000} for Co on Au(111),
one may model the Co adatom by an effective nondegenerate Anderson
impurity~\cite{Anderson61}, characterized by an effective energy
level $\epsilon_d$, an on-site repulsion $U$, and two hybridization
matrix elements $t_s$ and $t_b$ to the underlying surface and
bulk conduction electrons. In this manner, each of the Co atoms
forming the ellipse in the experiment of Manoharan
{\em et al.}~\cite{Mirage_00} acts as an Anderson impurity, as
does the adatom placed inside the ellipse.

\begin{figure}     
  \begin{center}    
\leavevmode    
        \epsfxsize=7cm       
         \epsfbox{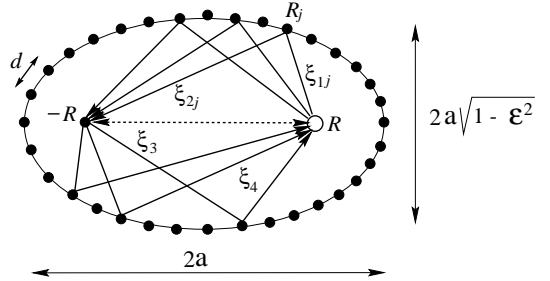}    
\end{center}  
\caption{An illustration of an elliptic quantum corral and
the classical trajectories associated with the quantum mirage.
The ellipse is characterized by a semimajor axis length $a$
and an eccentricity ${\cal E}$. The mean distance between
adjacent atoms forming the ellipse perimeter is $d$.}
\end{figure}       

Denoting the creation of a surface-state and a bulk conduction
electron by $c^{\dagger}_{\vec{k}\sigma}$ and
$a^{\dagger}_{\vec{q}\sigma}$, respectively (here $\vec{k}$
labels a two-dimensional surface vector while $\vec{q}$ is a
three-dimensional vector), we model the system by the Hamiltonian
${\cal H} = {\cal H}_{\rm surf} + {\cal H}_{\rm bulk}
+ \sum_{i=0}^{N} {\cal H}_{\rm imp}(\vec{R}_i)$, where
\begin{eqnarray}
{\cal H}_{\rm surf} = \sum_{\vec{k}\sigma} \epsilon_{\vec{k}}
                   c^{\dagger}_{\vec{k}\sigma} c_{\vec{k}\sigma}
~~\mbox{and}~~ {\cal H}_{\rm bulk} = \sum_{\vec{q}\sigma} E_{\vec{q}}
                   a^{\dagger}_{\vec{q}\sigma} a_{\vec{q}\sigma} \nonumber
\end{eqnarray}
describe the free surface  and bulk conduction bands, respectively, and
\begin{eqnarray}
{\cal H}_{imp}(\vec{R}_i) = &&
           \epsilon_d\sum_{\sigma}d^{\dagger}_{i\sigma}d_{i\sigma}
           + U d^{\dagger}_{i\uparrow}d_{i\uparrow}
           d^{\dagger}_{i\downarrow}d_{i\downarrow} \nonumber \\
&& + \sum_{\sigma} \left \{
           t_s d^{\dagger}_{i\sigma} \psi_{\sigma}(\vec{R}_i) +
           t_b d^{\dagger}_{i\sigma} \chi_{\sigma}(\vec{R}_i) +
           {\rm h.c.} \right \}
\nonumber
\end{eqnarray}
describes a Co adatom at point $\vec{R}_i$ on the surface.
Here $d^{\dagger}_{i\sigma}$ creates a localized Co electron
at site $\vec{R}_i$ ($i = 0$ for the inner adatom and
$i = 1, \cdots, N$ for the perimeter adatoms), while 
$\psi_{\sigma}(\vec{R}_i)$ and $\chi_{\sigma}(\vec{R}_i)$
annihilate, respectively, a surface and a bulk conduction
electron at site $\vec{R}_i$.
For simplicity, we have taken the different adatoms to be
identical, and neglected any momentum dependence of $t_s$
and $t_b$. In what
follows, we shall mainly be interested in the case where
the inner adatom is located at the left focus, i.e.,
$\vec{R}_0 = -\vec{R}$ in the notations of Fig.~1.

Consider now an STM tip placed directly above the surface point
$\vec{r}$. If the tip couples predominantly to the underlying
surface-state electrons at $\vec{r}$, then the differential
conductance for the current through the STM tip measures, up to thermal 
broadening, the local surface-electron density of states at
point $\vec{r}$, $\rho(\vec{r}, \epsilon)$. For an isolated Co
impurity, $\rho(\vec{r}, \epsilon)$ depends on the Kondo scattering
from the impurity as described, e.g., in Ref.~\onlinecite{SH_00}.
For the multiple-impurity configuration considered here there are
two main modifications: (i) There are multiple scattering off
the different Co adatoms; (ii) Intersite correlations alter the
Kondo scattering off each Co adatom. Due to the relatively large
distance between Co atoms ($\sim 10$\AA \ for neighboring atoms
on the ellipse perimeter), we expect the latter effect to be
small, and therefore neglect it hereafter.

Since we are mostly interested in the effect of the Co atom
placed inside the ellipse, we distinguish it from the other
Co atoms on the perimeter of the ellipse. Neglecting intersite
correlations, $\rho(\vec{r}, \epsilon)$ takes the form
$\rho(\vec{r}, \epsilon) =\bar{\rho}(\vec{r}, \epsilon)+
\delta \rho(\vec{r}, \epsilon)$, where 
\begin{eqnarray}
\bar{\rho} (\vec{r}, \epsilon) =  -\frac{1}{\pi}\mbox{Im} \{ G(\vec{r}, 
\vec{r}; \epsilon) \} 
\end{eqnarray}
is  the density of states of an empty ellipse (i.e.,
in the absence of the inner adatom), and
\begin{equation}
\delta \rho(\vec{r}, \epsilon)= -\frac{1}{\pi}
       {\rm Im} \{ t_s^2 G(\vec{r}, \vec{R}_0; \epsilon)
       G_{d}(\epsilon) G(\vec{R}_0, \vec{r}; \epsilon) \}
\label{mirage}
\end{equation}
is the additional contribution due to the extra Co atom at
$\vec{R}_0$. Here $G(\vec{r}, \vec{r}~\!'; \epsilon)$ 
is the retarded Green function of the surface electrons
for an empty ellipse, and $G_{d}(\epsilon)$ is the fully
dressed retarded Green function of the $d$ electrons of
the inner adatom. Note that in writing Eq.~(\ref{mirage})
we have assumed that the three-dimensional propagation
of bulk electrons between different Co sites on the surface
is small compared to the two-dimensional propagation of
the surface electrons. This assumption is quite reasonable
considering that the three-dimensional propagation near
the surface decays as $1/r^2$, compared to $1/\sqrt{r}$ for
the two-dimensional surface propagation~\cite{comment_on_3D}.

Experimentally, $\delta \rho(\vec{r}, \epsilon)$ is extracted
by first measuring the local density of states of the empty
ellipse, and then subtracting it from the measured density of
states with the extra Co atom. To compute
$\delta \rho(\vec{r}, \epsilon)$, one needs to evaluate
$G(\vec{r}, \vec{r}~\!'; \epsilon)$, which is our next goal.
To this end, we introduce the $N \times N$ matrix $g_{ij} \equiv
(1 - \delta_{i,j}) G_{0}(\vec{R}_i, \vec{R}_j)$, along
with the two vector quantities,
$v_i = G_{0}(\vec{r}, \vec{R}_i)$ and
$u_i = G_{0}(\vec{R}_i, \vec{r}~\!')$. Here 
$G_{0}(\vec{r}, \vec{r}~\!')$ 
is the free surface
Green function without the corral, and $i$ and $j$ run over
$1, \cdots , N$. Using these quantities, 
$G(\vec{r}, \vec{r}~\!';\epsilon)$ is compactly expressed as
\begin{equation}
G(\vec{r}, \vec{r}~\!' ;\epsilon) =
       G_{0}(\vec{r},\vec{r}~\!') +
       \sum_{i,j = 1}^{N} v_i \left [
       \frac{1}{ 1 - T g} T \right ]_{ij} u_j ,
\label{G}
\end{equation}
where $T(\epsilon) = t_{s}^2 G_{d}(\epsilon)$ is the
surface-to-surface component of the conduction-electron
scattering $T$-matrix at each Co site. Again, 
Eq.~(\ref{G}) omits the intersite correlations and the bulk
propagation between different Co sites. Finally, for
temperatures and energies below the Kondo
temperature, the Kondo part of $G_{d}(\epsilon)$
may be well approximated~\cite{Hewson} by the Lorentzian form
$Z_K/(\epsilon - \epsilon_F + i T_K)$, where $T_K$ is the
Kondo temperature, $\epsilon_F$ is the Fermi energy, and
\begin{equation}
Z_K = \frac{T_K}{\pi \rho_s t_s^2 + \pi \rho_b t_b^2}
\end{equation}
is the corresponding weight. Here $\rho_s$ and $\rho_b$ are the
surface and bulk density of states at the Fermi level.

A key parameter that enters the quantum mirage is the ratio
of scattering rates
\begin{equation}
t= \frac{ \pi \rho_s t_s^2}{\pi \rho_s t_s^2 + \pi \rho_b t_b^2} .
\end{equation}
Physically, $t$ represents the probability that a surface-state
electron impinging on a Co adatom will be scattered to a
surface-state electron rather than a bulk electron. Hence
$t$ is a measure of the in-elasticity of the scattering of surface
waves from the Co impurities. In the theory
of Heller {\em et al.}~\cite{Heller} for the standing waves formed
in a quantum corral, $t$ is found to be $1/2$. Hereafter we shall 
use the same value for $t$.

{\narrowtext   
\begin{figure}     
  \begin{center}    
\leavevmode    
        \epsfxsize=4.2cm       
         \epsfbox{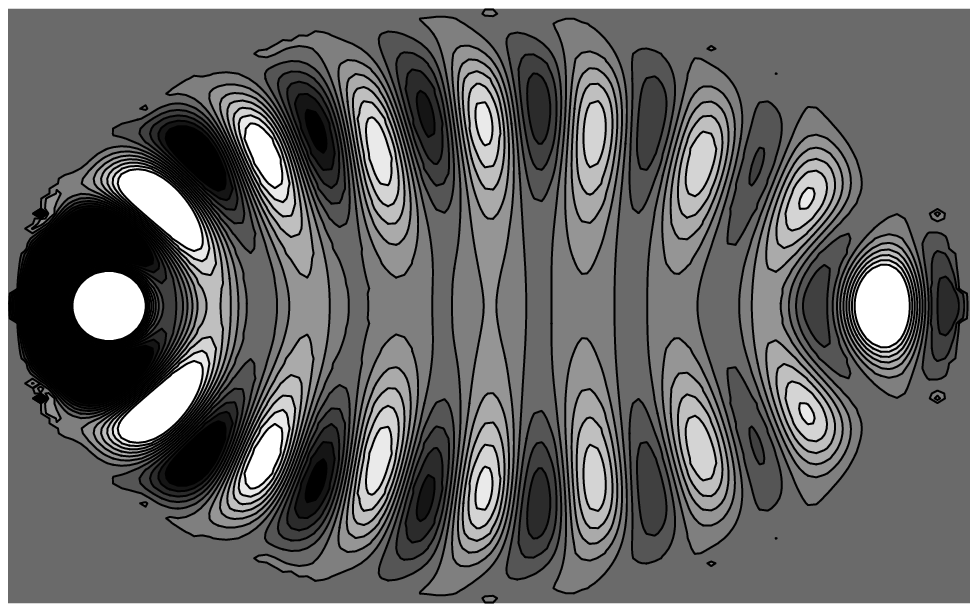}    
\leavevmode 
        \epsfxsize=4.2cm       
         \epsfbox{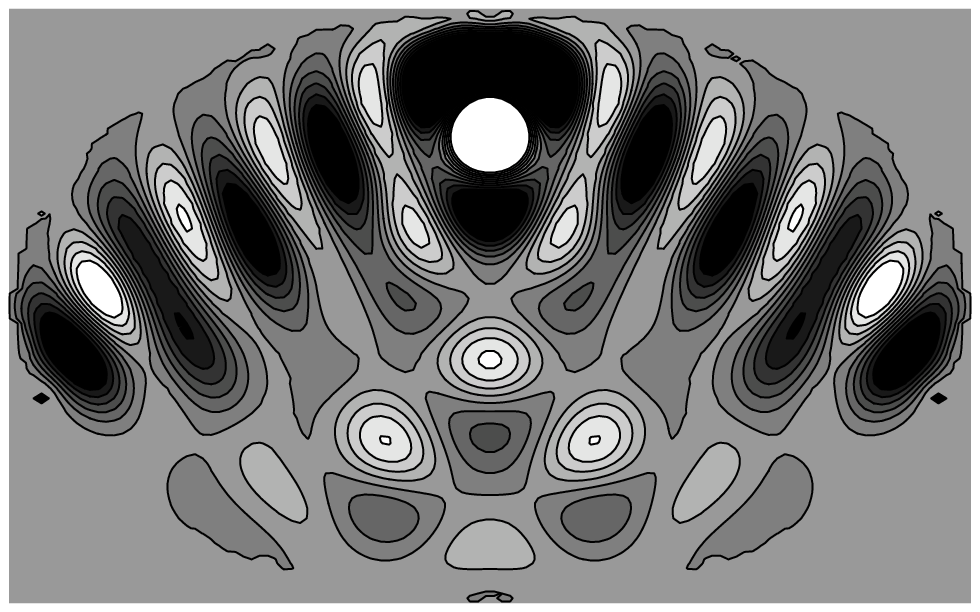}
 
\end{center}  
\vspace{-0.7cm}
  \begin{center}    
\leavevmode    
        \epsfxsize=4.2cm       
         \epsfbox{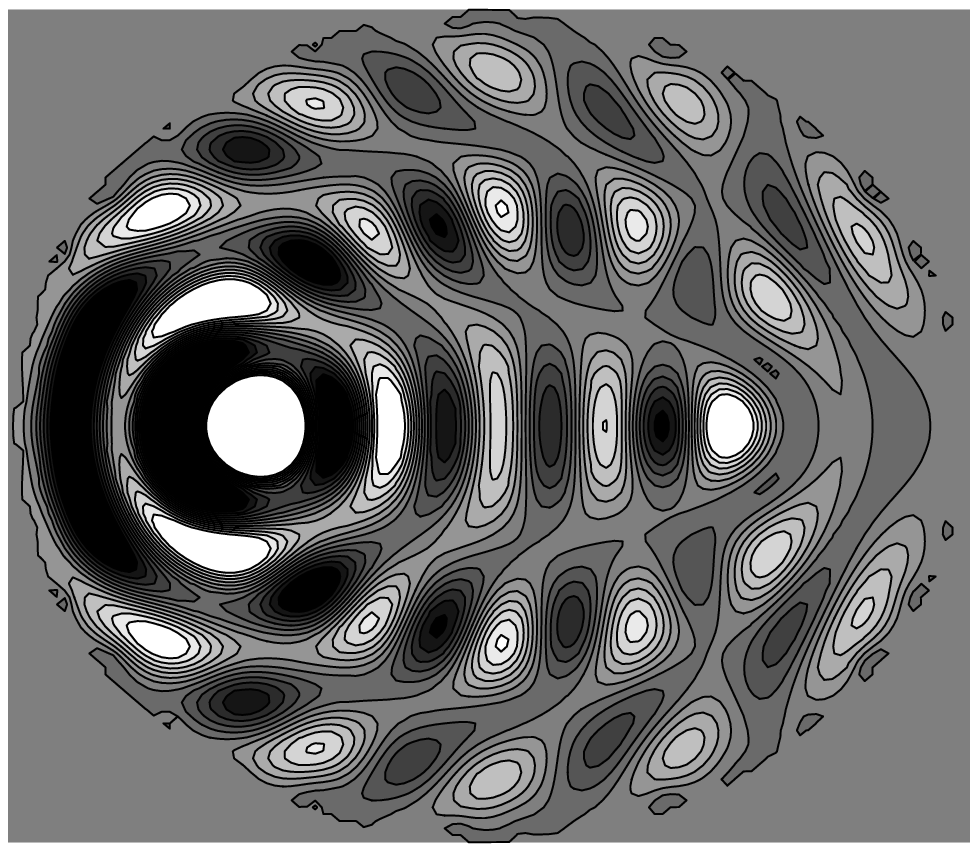}    
\leavevmode 
        \epsfxsize=4.2cm       
         \epsfbox{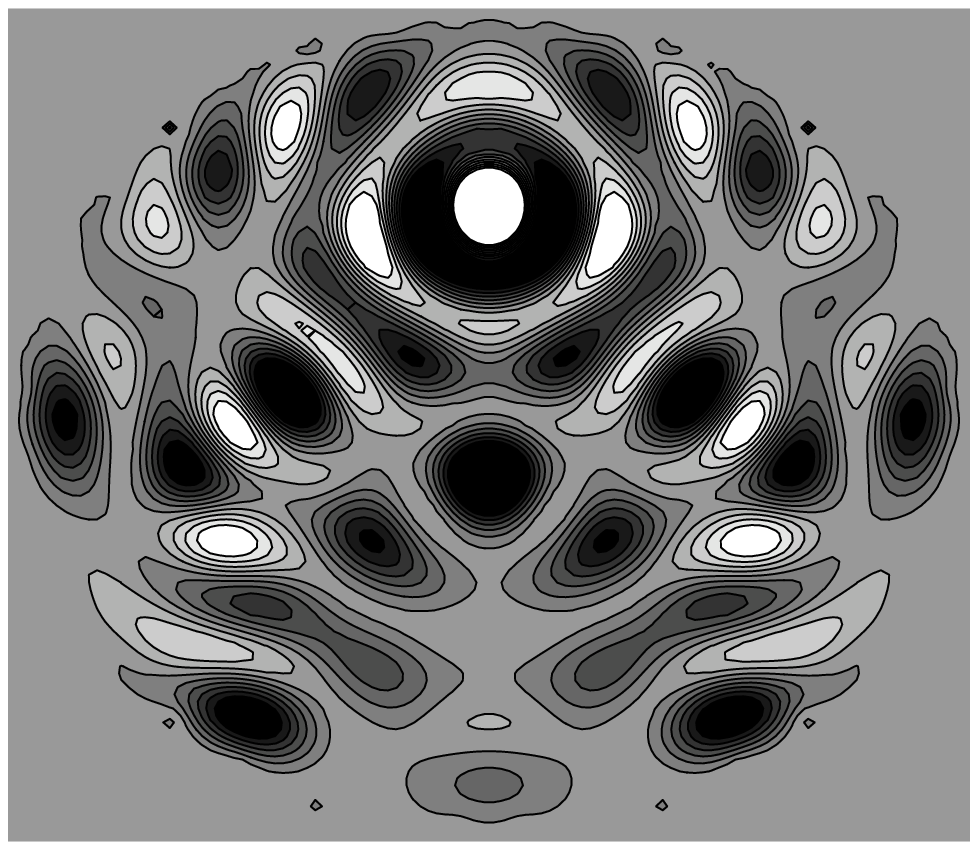}
 
\end{center}  
    
\caption{Contour plot of 
$\delta \rho(\vec{r}, \epsilon_F)$ at the 
Fermi level assuming $t=1/2$. The upper
panels correspond to an ellipse with eccentricity
${\cal E}=0.786$ and 34 adatoms, while the lower panels
to an ellipse with ${\cal E}=0.5$
and 36 adatoms. For both ellipses, a strong signal
appears at the right focus of the ellipse when an additional adatom 
has been placed at the left focus. This quantum mirage disappears
when the additional adatom is places off the focus, as shown in
the right two panels.
}              
\end{figure}       
}   

In Fig.~2 we depict $\delta \rho(\vec{r}, \epsilon_F)$ for
various configurations of the Co atoms, as measured by
Manoharan {\em et al.}~\cite{Mirage_00}. The upper
panels correspond to an ellipse with eccentricity
${\cal E}=0.786$ and 34 adatoms (ellipse {\bf b} of
Ref.~\onlinecite{Mirage_00}), while the lower panels
correspond to an ellipse with eccentricity ${\cal E}=0.5$
and 36 adatoms (ellipse {\bf a} in Ref.~\onlinecite{Mirage_00}). 
The quantum mirage is clearly seen in each of the left two panels,
where an additional adatom has been placed at the left focus of
the ellipse. For both ellipses, there is a strong signal in the
tunneling density of states right above the right focus, in accordance
with the experimental data. By contrast, the quantum mirage disappears
when the additional adatom is places off the focus, as shown in
the right two panels. These results are in good agreement with
the experimental measurements, reproducing even fine details of
the experimental patterns. 

The results of Fig.~2 were obtained from Eqs.~(\ref{mirage})
and (\ref{G}), by setting $t=1/2$ and approximating
$G_0$ with the free two-dimensional Green function:
\begin{eqnarray}
G_0(\vec{r}, \vec{r}~\!') =
        -i \pi \rho_s H_0^{(1)}(k|\vec{r}-\vec{r}~\!'|).
\end{eqnarray}
Here $k$ is the wave number, and $H_0^{(1)}(x)$ is the Hankel
function of zeroth order, which for $x\gg 1$ takes
the asymptotic form 
\begin{eqnarray}
H_0^{(1)}(x)\simeq \sqrt{\frac{2}{\pi x}} \exp 
\left( i x - i \frac{\pi}{4} \right).
\end{eqnarray}

The good agreement between our calculations with $t=1/2$ and
the experimental data indicates that the number of scattering
events that a particle undergoes before leaving the surface
bounded by the ellipse is small. This suggests the possibility
of calculating the quantum mirage at the right focus at the Fermi 
level, $\delta \rho(\vec{R}, \epsilon_F)$, using perturbation theory
in $t$. Thus, to linear order in $t$, the Green function
between the left and the right foci is given by
\begin{eqnarray}
G(-\vec{R},\vec{R};\epsilon_F) \simeq
       G_0(-\vec{R},\vec{R})+G_1(-\vec{R},\vec{R}) +\cdots ,
\label{expa}
\end{eqnarray}
where
\begin{eqnarray}
G_0(-\vec{R},\vec{R}) \simeq -i \rho_s
     \sqrt{\frac{\pi}{ {\cal E} k_F a }}
     e^{ i 2 {\cal E} k_F a  - i \frac{\pi}{4}}
\end{eqnarray}
is the contribution of the direct path
connecting the two foci (illustrated by the dashed line in
Fig.~1), and 
\begin{eqnarray}
G_1(-\vec{R},\vec{R})= \frac{t}{i \pi \rho_s}
\sum_{j=1}^{N} G_0(-\vec{R}, \vec{R}_j)
G_0(\vec{R}_j, \vec{R})
\end{eqnarray}
comes form all trajectories which scatter from a single
Co adatom on the ellipse perimeter (solid lines in Fig.~1).

Conventionally, the second term $G_1$ is smaller than $G_0$ for
several reasons. First, each scattering event is inelastic,
and therefore introduces a reduction factor of $t$. Second,
the scattered orbits are longer. Third, the various orbits
have generally different lengths, and therefore their
corresponding phases add up incoherently.
The situation is quite different in the present case.
Due to the defining property of the ellipse, the length
of all scattered orbits between the two foci is precisely
the same, and hence their contributions add up coherently.
Consequently, the second term in Eq.~(\ref{expa}) takes
the form
\begin{eqnarray}
G_1(-\vec{R}, \vec{R}) \simeq  i \rho_s \sum_{j=1}^N
      \frac{2 t}{ k_F \sqrt{ \xi_{1,j} \xi_{2,j}} }
      e^{ i 2 k_F a - i \frac{\pi}{2}},
\label{G1}
\end{eqnarray}
where $2a$ is the length of each orbit, and $\xi_{1,j}$ and
$\xi_{2,j}$ are the distances between the impurity at
$\vec{R}_j$ and the right and left foci, respectively
(see Fig.~1). Finally, we approximate the sum in
Eq.~(\ref{G1}) by an integral:
\begin{eqnarray}
\sum_{j=1}^N \frac{1}{\sqrt{ \xi_{1,j} \xi_{2,j}}}
       \simeq \frac{1}{d} \oint ds 
       \frac{1}{\sqrt{ \xi_1(s) \xi_2(s)}} ,
\label{continuum}
\end{eqnarray}
where $s$ denotes the coordinate along the ellipse contour,
and $d$ is the mean distance between adjacent adatoms. The
result of the integral is independent of the eccentricity
of the ellipse, and is simply $2 \pi$. Thus, the contribution
of orbits scattered from a single perimeter adatom is
\begin{eqnarray}
G_1(-\vec{R}, \vec{R}) \simeq 
       \rho_s \frac{4 \pi t}{ k_F d} 
      e^{ i 2 k_F a }.
\label{fg1}
\end{eqnarray} 

Comparing $G_0(-\vec{R}, \vec{R})$ and 
$G_1(-\vec{R}, \vec{R})$ at the Fermi energy, one sees
that the leading contribution to the quantum mirage
comes from $G_1$ \cite{comment}, provided
\begin{equation}
    \frac{d}{{\cal E} a} \ll \frac{ 16 \pi t^2}{k_F d}.
\label{inq}
\end{equation} 
Substituting the experimental parameters, $d \simeq 10$\AA, 
$a\simeq 70$\AA, and $k_F\simeq 1/4.75$\AA$^{-1}$, and setting
$t=1/2$, it is straightforward to verify that Eq.~(\ref{inq})
holds for all ellipses with eccentricity $0.05 < {\cal E} < 1$.
Furthermore, Eq.~(\ref{inq}) is always satisfied for sufficiently
large ellipses, provided the mean distance between adjacent
adatoms is kept fixed ($a\gg d$).

Neglecting the contribution of the direct path,
$G_0(-\vec{R}, \vec{R})$, and using Eqs.~(\ref{mirage})
and (\ref{fg1}) with $\vec{r} = \vec{R}$, the resulting
local density of states at the Fermi energy takes the form
\begin{equation} 
\delta \rho (\vec{R}, \epsilon_F) \simeq
      \rho_s \frac{ 16 t^3}{ (k_F d)^2} 
      \cos ( 4 k_F a) .
\label{central}
\end{equation}
The main feature of the above result is the robustness of the
quantum mirage:
As long as condition (\ref{inq}) is satisfied, the amplitude
of the mirage is independent of the size of the ellipse, 
$a$, and its eccentricity, ${\cal E}$. Rather, the amplitude is
determined by $t$, which characterizes the in-elasticity
of the scattering of surface waves from adatoms, and the
dimensionless mean distance between adjacent adatoms along
the ellipse, $k_Fd$. The oscillations of the mirage as function
of $a$ are indeed periodic with a period of $\lambda_F/4$,
as seen experimentally~\cite{Mirage_00}.

Next we consider the effect of a weak uniform magnetic
field, $B$, applied perpendicular to the surface. As
shown below, the quantum mirage experiences a distinctive
oscillatory behavior as a function of the magnetic field,
which depends on the size and the shape of the ellipse. Here we assume
that the ellipse is sufficiently large and that the magnetic
field is sufficiently weak so that (i) Zeeman splitting of
the Kondo resonance can be neglected, and (ii) the
cyclotron radius of the conduction electrons is much
larger than the ellipse size.

Under these circumstances, the main effect of the magnetic
field is to introduce an additional Aharonon-Bohm phase to
the contribution of each path. This phase is the magnetic
flux encircled by the orbit, measured in units of the
quantum flux, $\phi_0= h c/e$. Here $h$ is Planck's constant,
$c$ is the velocity of light, and $e$ is the electron
charge. To compute $G_1(-\vec{R},\vec{R})$ we fix the
gauge by calculating the flux encircled by the path
which goes from $\vec{R}$ to $\vec{R}_j$, to $-\vec{R}$,
and then back to $\vec{R}$ along the semimajor axis of the
ellipse. Accordingly, the sum of Eq.~(\ref{G1}) is modified to
$\sum_j (\xi_{1,j}\xi_{2,j})^{-1/2}
e^{ i 2 \pi \varphi_j/\phi_0}$,
where $\varphi_j$ is the flux of each trajectory. Using the
continuum approximation of Eq.~(\ref{continuum}) this sum
gives $2 \pi J_0( 2 {\cal E} {\cal A} B/\phi_0)$,
where $J_0(x)$ is the Bessel function of zeroth order,
and ${\cal A}= \pi a^2\sqrt{1-{\cal E}^2}$ is the area of
the ellipse. The quantum mirage is, thus, modified
according to
\begin{eqnarray}
\delta \rho (B)\simeq \delta \rho (0)
      J_0^2\left(2 {\cal E} \frac{{\cal A} B}{\phi_0}\right), 
\end{eqnarray}
where $ \delta \rho (0)$ is the zero-field result of
Eq.~(\ref{central}). Note that, for a given $a$, the
sensitivity to a magnetic field is largest for
${\cal E} = 1/\sqrt{2}$.

Finally, we discuss the role of imperfections in the ellipse.
Clearly, a combined effect of many coherent trajectories is very
sensitive to imperfections and dephasing. At 4K, dephasing effects
due to electron-electron and electron-phonon interactions are
negligible over distances of the order of hundreds of angstroms.
In what follows we show that effect of imperfections is small too.

The main source of imperfections in the ellipse comes from the
position of the adatoms forming the ellipse. These are constrained 
to sit on a triangular lattice imposed by the underlying Cu(111)
surface. Consequently, the lengths of the orbits contributing to
$\delta \rho$ are not those of an ideal ellipse. To estimate
the effect of the deviations, we consider a random distribution
of the trajectory lengths, and average over the distribution.
We first notice that each contribution to $\delta\rho$ is
composed of 4 segments: $\xi_1$, $\xi_2$, $\xi_3$, and $\xi_4$,
as illustrated in Fig.~1. Each of these segments
is regarded as an independent random variable uniformly distributed 
in the range $(\bar{\xi}-b/2,\bar{\xi}-b/2)$, where $\bar{\xi}$ 
is the exact distance from the focus to the ellipse boundary, and
$b$ is the triangular lattice spacing. The total length
$\eta=\sum_{n=1,4}\xi_n$ is, therefore, approximately 
a Gaussian random variable with mean $4a$ and variance $b^2/3$.
Averaging the cosine term in Eq.~(\ref{central}),
$\langle \cos ( k_F \eta) \rangle$, the disorder-averaged value
of the quantum mirage is reduced by a factor of approximately
$Q=\exp (-k_F^2 b^2 /6)$. 
Substituting the experimental values $k_F=1/4.75$\AA$^{-1}$ 
and $b=2.55$\AA, one finds $Q\approx 0.95$, 
meaning that the effect of imperfections is negligible.

In conclusion, in this Letter we have studied the phenomenon of
quantum mirage, and clarified its relation to the classical
orbits of a particle in an ellipse. Our theory also predicts
a distinctive behavior of the quantum mirage in the presence 
of a perpendicular magnetic field, which could be tested
experimentally. Finally our approach clearly shows that 
the phenomenon of quantum mirage is not unique for 
magnetic adatoms. It will also appear for nonmagnetic atoms
(on the ellipse perimeter or at its focus)
with strong scattering, e.g. adatoms with resonant
tunneling states at the Fermi level.  

A.S. is grateful to Hari Manoharan for useful discussions.
This research was supported by the Israel science foundation
founded by The Israel Academy of Science and Humanities, and
by Grant No.~9800065 from the USA-Israel
Binational Science Foundation (BSF).


\begin{thebibliography}{9}
\bibitem{comment} One may be concerned about higher order terms, such
    as $G_2$, $G_3$, etc. These, however, are small, as they represent
    mainly non-classical paths, whose contributions add up incoherently.
\end{thebibliography}
\end{document}